\documentclass[manuscript]{acmart}

\usepackage{bm}

\usepackage{amsmath}
\usepackage{subfigure}
\usepackage{amsthm}
\usepackage{graphicx}
\usepackage{float}
\usepackage{threeparttable}
\usepackage{multirow}
\usepackage{capt-of}
\AtBeginDocument{%
  }

\copyrightyear{2023} 
\acmYear{2023} 
\setcopyright{acmlicensed}\acmConference[RecSys '23]{Seventeenth ACM Conference on Recommender Systems}{September 18--22, 2023}{Singapore, Singapore}
\acmBooktitle{Seventeenth ACM Conference on Recommender Systems (RecSys '23), September 18--22, 2023, Singapore, Singapore}
\acmPrice{15.00}
\acmDOI{10.1145/3604915.3608813}
\acmISBN{979-8-4007-0241-9/23/09}

\begin{document}

\title{ADRNet: A Generalized Collaborative Filtering Framework Combining Clinical and Non-Clinical Data for Adverse Drug Reaction Prediction}

\author{Haoxuan Li}
\email{hxli@stu.pku.edu.cn}
\affiliation{%
  \institution{Peking University}
  \city{Beijing}
  \country{China}
}

\author{Taojun Hu}
\affiliation{%
  \institution{Peking University}
  \city{Beijing}
  \country{China}}
\email{larst@affiliation.org}

\author{Zetong Xiong}
\affiliation{%
  \institution{Yale University}
  \city{New Haven}
  \country{USA}
}

\author{Chunyuan Zheng}
\affiliation{%
 \institution{University of California, San Diego}
 \city{San Diego}
 \country{USA}}

\author{Fuli Feng}
\affiliation{%
  \institution{University of Science and Technology of China	}
  \city{Hefei}
  \country{China}}

\author{Xiangnan He}
\affiliation{%
  \institution{University of Science and Technology of China	}
  \streetaddress{8600 Datapoint Drive}
  \city{Hefei}
  \country{China}
  }

\author{Xiao-Hua Zhou}
\affiliation{%
  \institution{Peking University}
  \city{Beijing}
  \country{China}}
  \authornote{Xiao-Hua Zhou is the corresponding author.}
\email{jsmith@affiliation.org}


\renewcommand{\shortauthors}{Haoxuan Li et al.}

\begin{abstract}
Adverse drug reaction (ADR) prediction plays a crucial role in both health care and drug discovery for reducing patient mortality and enhancing drug safety. Recently, many studies have been devoted to effectively predict the drug-ADRs incidence rates. However, these methods either did not effectively utilize non-clinical data, i.e., physical, chemical, and biological information about the drug, or did little to establish a link between content-based and pure collaborative filtering during the training phase. In this paper, we first formulate the prediction of multi-label ADRs as a drug-ADR collaborative filtering problem, and to the best of our knowledge, this is the first work to provide extensive benchmark results of previous collaborative filtering methods on two large publicly available clinical datasets. Then, by exploiting the easy accessible drug characteristics from non-clinical data, we propose ADRNet, a generalized
collaborative filtering framework combining clinical and non-clinical data for drug-ADR prediction. Specifically, ADRNet has a shallow collaborative filtering module and a deep drug representation module, which can exploit the high-dimensional drug descriptors to further  guide the learning of low-dimensional ADR latent embeddings, which incorporates both the benefits of collaborative filtering and representation learning. Extensive experiments are conducted on two publicly available real-world drug-ADR clinical datasets and two non-clinical datasets to demonstrate the accuracy and efficiency of the proposed ADRNet. The code is available at \url{https://github.com/haoxuanli-pku/ADRnet}.
\end{abstract}





\maketitle

\section{Introduction}
Adverse drug reactions (ADRs, also known as side effects) refer to harmful effects produced by drugs that are detrimental to the patient's treatment in normal therapy~\cite{edwards2000adverse,karimi2015text}. Until now, thousands types of ADRs have been reported, many of which have led to serious unwanted and harmful consequences~\cite{wu2010ten}. Timely and effective warning of ADRs can help regulate as well as guide the production of drugs with limited side effects. Therefore, the accurate prediction of ADRs is meaningful in both health care~\cite{houck2008understanding} and drug discovery~\cite{arrowsmith2013trial} for improving drug safety and reducing patient mortality.

\begin{figure}[t]
    \centering
\includegraphics[scale=0.37]{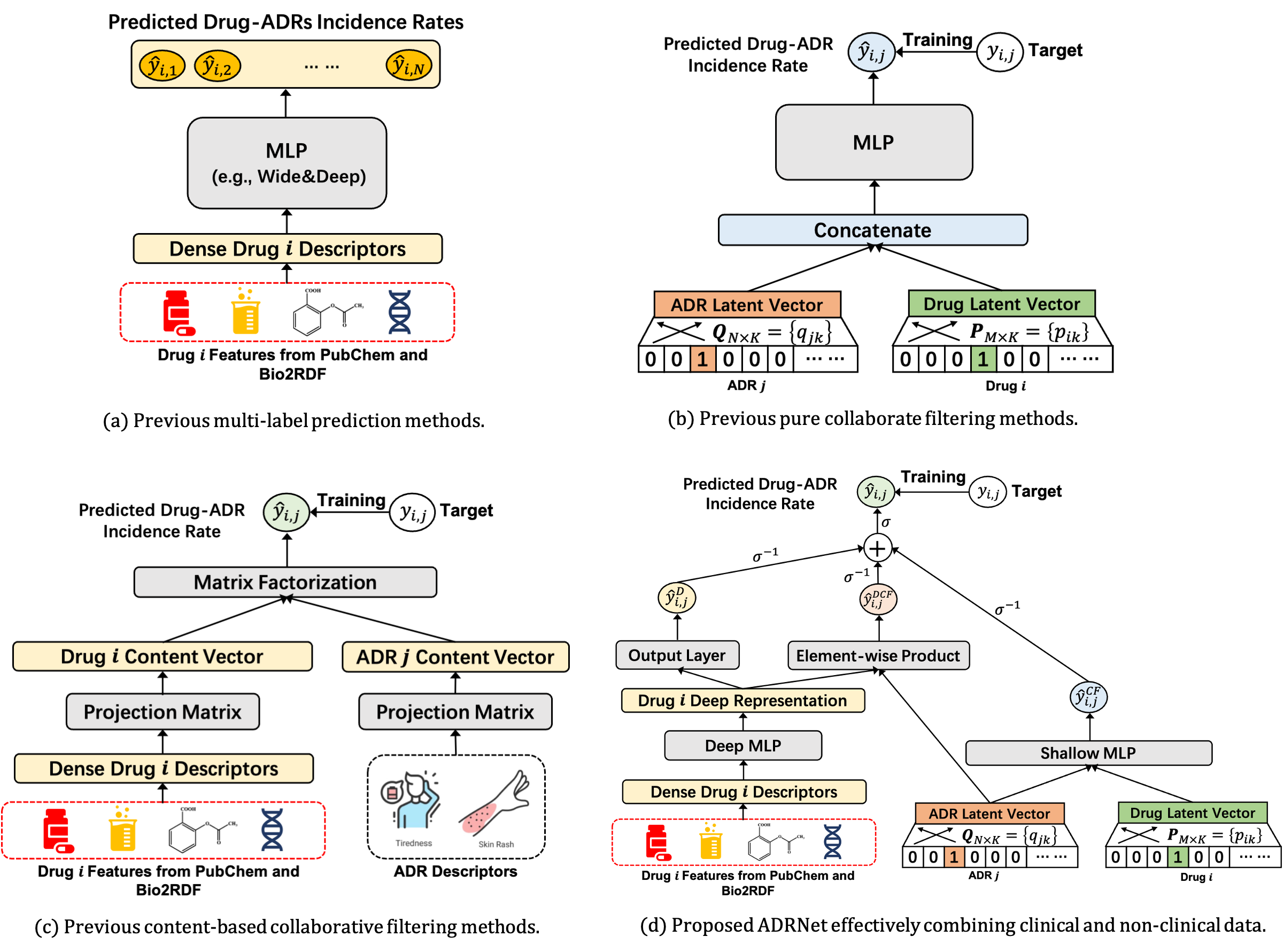}
    \vspace{-0.55cm}
    \caption{Comparisons of (a) previous multi-label prediction methods; (b) previous pure collaborative filtering methods; (c) previous content-based collaborative filtering methods; (d) proposed ADRNet as a generalized collaborative filtering framework combining clinical and non-clinical data for drug-ADR prediction, contains (i) a deep drug representation module; (ii) a shallow latent collaborative filtering module; and (iii) a drug collaborative filtering module.}
    \label{fig:1} 
    \vspace{-0.4cm}
\end{figure}

Many previous approaches are based on association rule mining and statistical significance tests~\cite{agresti1992survey} to identify the significance of drug-ADR associations~\cite{chen2001prediction,vedani2005silico,fliri2005analysis,tatonetti2009predicting,hammann2010prediction,montastruc2011benefits}, but with limited performance on prediction tasks.  An alternative class of deep learning-based methods uses pharmacovigilance networks~\cite{cami2011predicting,davazdahemami2018chronological}, ensemble approaches~\cite{jahid2013ensemble}, and deep neural models~\cite{wang2019detecting,sedhain2015autorec}, and has achieved promising predictive performance for specific adverse reactions, as shown in Figure \ref{fig:1}(a). However, as drugs may cause multiple ADRs simultaneously, the direct repeat use of the single-label prediction methods may lead to time-consuming and suboptimal prediction performance.

To tackle this issue, collaborative filtering methods can efficiently and accurately predict the relationship between drugs and multiple ADRs. The underlying motivation is that drugs with similar interactions in clinical data tend to have similar ADRs. For example, matrix factorization (MF)~\cite{koren2009matrix,pauwels2011predicting} and neural matrix factorization (NMF)~\cite{he2017neural} first learn a latent embedding for each drug and ADR, and then perform drug-ADR prediction. However, as shown in Figure \ref{fig:1}(b), the pure collaborative filtering methods ignore the non-clinical data (e.g., drug descriptors~\cite{kim2016pubchem}), which are readily available and can significantly improve the drug-ADR prediction performance. Recently, content-based collaborative filtering methods have achieved impressive performance in recommendation, as shown in Figure \ref{fig:1}(c), by incorporating user-side and item-side features, the prediction performance can be further improved. \emph{However, unlike traditional recommender systems where users and items have symmetric information and status, data collected for drugs and ADRs usually contain non-clinical drug characteristics but lack of ADR characteristics (e.g., cough and fever), due to the inconvenience of the ADR characteristics to be obtained and processed, which poses a great challenge for developing effective drug-ADR predictions using both clinical and non-clinical data.}

In this paper, we first formulate the prediction of multi-label ADRs as a drug-ADR collaborative filtering problem, and to the best of our knowledge, this is the first study to provide extensive benchmark results of previous collaborative filtering methods on two large publicly available clinical datasets. Next, by noting the easy accessibility of drug characteristics and the asymmetric information of drugs and ADRs that differs from previous collaborative filtering, we propose ADRNet, a generalized collaborative filtering framework combining
clinical and non-clinical data for drug-ADR prediction. Specifically,ADRNet has a shallow collaborative filtering module and a deep drug representation module, which can exploit the high-dimensional drug descriptors to further  guide the learning of low-dimensional ADR latent embeddings, which incorporates both the benefits of collaborative filtering and representation learning. 

Notably, the proposed ADRNet effectively exploits
 both the advantages of deep content-based and latent collaborative filtering-based approaches. On one hand, compared with the deep drug feature-based networks, our approach has fewer parameters and is convenient. Previous studies mapped drugs to ADR space~\cite{dimitri2017drugclust,xiao2017adverse,lorberbaum2016integrative} or to metabolic reaction space~\cite{shaked2016metabolic} and applied flux variability analysis to represent drug-protein/gene interactions~\cite{mahadevan2003effects}, whereas our approach further utilize the learned ADR latent embedding from collaborative filtering to guide the deep representations learning of drugs. On the other hand, compared with the latent collaborative filtering-based approaches, the proposed approach has the clear advantage of exploiting the chemical, physical and biological characteristics of drugs contained in non-clinical data~\cite{belleau2008bio2rdf} to help reveal the mechanisms of drug-ADR and improve the prediction performance.

The contributions of this paper are summarized as follows.
\begin{itemize}
    \item[$\bullet$] We formulate the prediction of multi-label ADRs as a drug-ADR collaborative filtering problem, and to the best of our knowledge, this is the first study to provide extensive benchmark results of previous collaborative filtering methods on two large publicly available clinical datasets.
    \item[$\bullet$] By exploiting the easy accessible drug characteristics from non-clinical data, we propose ADRNet, a generalized collaborative filtering framework combining
clinical and non-clinical data for drug-ADR prediction.
    \item[$\bullet$] We further propose a sharing mechanism for ADR latent vectors that is easily applicable to previous pure collaborative filtering methods, and a learning approach to jointly train a deep drug representation and a shallow collaborative filtering network to better trade-off the predictions of each sub-network.
    \item[$\bullet$] Extensive experiments are conducted on two publicly available real-world drug-ADR clinical datasets and two non-clinical datasets to demonstrate the accuracy and efficiency of the proposed ADRNet.
\end{itemize}

\section{Preliminaries and Problem Setup}
Let $M$ and $N$ be the number of drugs and ADRs, respectively. Based on the clinical data containing drug-ADR records from electronic health or adverse reporting systems~\cite{liu2012large,banda2016curated}, we define the drug-ADR interaction matrix $\mathbf{Y} \in \mathbb{R}^{M \times N}$ as
\[
y_{i,j}= \begin{cases}1, & \text { if interaction ( drug } i, \text { ADR } j \text { ) is observed; } \\ 0, & \text { otherwise. }\end{cases}
\]
The multi-label ADR prediction problem is formalized as a collaborative filtering problem to accurately predict drug-ADR incidence rate. Specifically, the model-based approach is abstracted as learning $\hat{y}_{i,j}=f(i,j \mid \Theta)$, where $\hat{y}_{i,j}$ is the predicted incidence rate of drug $i$ to ADR $j$, and $f$ is a parametric prediction model, $\Theta$ is the model parameters.

To accurately estimate parameters $\Theta$ in pure collaborative filtering, we associate each drug $i$ and ADR $j$ with a real-valued vector of hidden features, denoted as $\mathbf{p}_i$ and $\mathbf{q}_j$, respectively. 
Then many standard collaborative filtering methods can be adopted, e.g., matrix factorization (MF)~\cite{koren2009matrix,pauwels2011predicting} using the inner product of $\mathbf{p}_i$ and $\mathbf{q}_j$ for drug-ADR interaction prediction as
$\hat{y}_{i,j}=f\left(i,j \mid \mathbf{p}_i, \mathbf{q}_j\right)=\mathbf{p}_i^\top \mathbf{q}_j=\sum_{k=1}^K p_{i k} q_{j k},$
where $K$ denotes the dimension of the latent space. 
More generally, the backbone can be substituted by neural networks
$
\hat{y}_{i,j}=f\left(\mathbf{P}, \mathbf{Q}\mid  \Theta\right),
$
where $\mathbf{P} \in \mathbb{R}^{M \times K}$ and $\mathbf{Q} \in \mathbb{R}^{N \times K}$, denoting the latent factor matrix for drugs and ADRs, respectively. However, these methods do not exploit drug features in easily available non-clinical data, leading to suboptimal performance in practice.

Content-based collaborative filtering further utilizes non-clinical data that contain chemical, physical and biological information about the drugs, which can be beneficial for the prediction of drug-ADR relations~\cite{testa1991concept,consonni2009handbook,grisoni2018molecular,yamanishi2012drug}. Specifically, drug descriptors in the non-clinical data can be categorized into two types: physical, chemical descriptors
(PC-descriptors), and
biological descriptors
(BIO-descriptors)~\cite{nguyen2021survey}. To enhance the prediction accuracy of drug-ADR, we use PubChem~\cite{kim2016pubchem} as the PC-descriptors and Bio2RDF~\cite{belleau2008bio2rdf} as the BIO-descriptors for drugs, and denote them with $\mathbf{x}^{PC}_i$ and $\mathbf{x}^{BIO}_i$, respectively. 

Unfortunately, unlike drug descriptors, ADRs in non-clinical data always appear in a descriptive manner (e.g., cough and fever), and these features are difficult to be processed and utilized directly to improve the prediction performance. Such drug-ADR information asymmetry poses a serious challenge to the previous collaborative filtering approaches. In order to simultaneously take the advantage of the latent collaborative filtering and exploit the informative drug descriptors, we propose a general asymmetric collaborative filtering framework for accurate drug-ADR prediction next.
\section{Proposed ADRNet: A Generalized Collaborative Filtering Framework}
In this section, we propose ADRNet, a generalized
collaborative filtering framework combining clinical and non-clinical data for drug-ADR prediction. By learning the latent embeddings from clinical data and leveraging high-dimensional drug descriptors from non-clinical data, the proposed method combines the benefits of collaborative filtering and deep representation learning. We introduce the structure of ADRNet as shown in Figure \ref{fig:1}(d) in the following subsections, including (i) a deep drug representation module in Section \ref{4.1}; (ii) a shallow latent collaborative filtering module in Section \ref{4.2}; and (iii) a drug collaborative filtering module in Section \ref{4.3}. We discuss the training of ADRNet and summarize its advantages in Section \ref{4.4}.
\subsection{Deep Drug Representation Module}\label{4.1}
The deep drug representation module first concatenates and maps sparse and high-dimensional drug PC-descriptors and BIO-descriptors to a low-dimensional and dense real-valued vector, also known as the drug embedding vector. These embedding vectors are randomly initialized, then fed into a feed-forward neural network, and obtain a drug representation \emph{with the same dimensions} as the drug/ADR latent space in latent collaborative filtering. Given a learned drug representation, as in Figure \ref{fig:1}(d), the representation is fed into a fully connected (FC) layer to obtain drug-level predictions $\hat y_{i, j}^D=\hat y_{i}^D$. In addition, this representation is shared to make predictions together with the ADR latent vector obtained from the collaborative filtering part. Specifically, the deep drug representation layers under our ADRNet framework are defined as 
\begin{align*}
&\mathbf{z}_1^{D} =\phi_1(\mathbf{x}^{PC}_i, \mathbf{x}^{BIO}_i)=\left[\begin{array}{l}
\mathbf{x}^{PC}_i \\
\mathbf{x}^{BIO}_i
\end{array}\right], \quad 
\mathbf{z}_2^{D}=\phi_2^{D}(\mathbf{z}_1^{D})  =a_2\left((\mathbf{W}_2^{D})^\top \mathbf{z}_1^{D}+\mathbf{b}_2^{D}\right), \quad
 \ldots, \\ 
&\mathbf{z}_L^{D}=\phi_L^{D}(\mathbf{z}_{L-1}^{D})  =a_L\left((\mathbf{W}_L^{D})^\top \mathbf{z}_{L-1}^{D}+\mathbf{b}_L^{D}\right), \quad
\hat{y}_{i, j}^{D}= \hat{y}_{i}^{D}  =\sigma\left((\mathbf{h}^{D})^\top \mathbf{z}_L^{D}\right),
\end{align*}
where $\mathbf{W}_l, \mathbf{b}_l$, and $a_l(\cdot)$ denote the weight matrix, bias vector, and activation function of the $l$-th layer's perceptron, and $\sigma(\cdot)$ denote the sigmoid function.

\vspace{-3pt}
\subsection{Shallow Latent Collaborative Filtering Module}\label{4.2}
The shallow latent collaborative filtering module concatenates the drug and ADR latent vectors and performs pure collaborative filtering using a feed-forward neural network, which is able to capture the non-linearity of interactions between $\mathbf{p}_i$ and $\mathbf{q}_j$. Compared with the drug representation module, the collaborative filtering module does not directly use the drug features, and therefore has shallower layers. However, in contrast to neural collaborative filtering (NCF)~\cite{he2017neural}, the learned drug representations can further guide the training of ADR latent vectors in Figure \ref{fig:1}(d), which provides a stronger supervised signal for latent vectors compared with collaborative filtering using only clinical interactions, and thus leads to more accurate predictions. Another advantage is induced from the joint learning of the deep network in Section \ref{4.1} and the shallow collaborative filtering. As suggested by the previous studies, jointly learning a shallow and a deep network can further improve the model's "memory" and "generalization" abilities~\cite{cheng2016wide,wang2017deep}. Specifically, the "memory" ability means that the model directly learns and utilizes the co-occurrence frequency of drugs and ADRs from historical data, and the "generalization" ability refers to letting the model mine the drug-ADR associations, expecting the model to be able to accurately predict sparse or even unobserved drug or ADR features. 
Formally, the shallow latent collaborative filtering module is formulated as
\begin{align*}
&\mathbf{z}_1^{CF} =\phi_1(\mathbf{p}_i, \mathbf{q}_j)=\left[\begin{array}{l}
\mathbf{p}_i \\
\mathbf{q}_j
\end{array}\right],\quad \mathbf{z}_2^{CF} =\phi_2^{CF}(\mathbf{z}_1^{CF})=a_2\left((\mathbf{W}_2^{CF})^\top \mathbf{z}_1^{CF}+\mathbf{b}_2^{CF}\right), \quad
 \ldots, \\
&\mathbf{z}_S^{CF}=\phi_S^{CF}(\mathbf{z}_{S-1}^{CF}) =a_S\left((\mathbf{W}_S^{CF})^\top \mathbf{z}_{S-1}^{CF}+\mathbf{b}_S^{CF}\right), \quad
\hat{y}_{i,j}^{CF} =\sigma\left((\mathbf{h}^{CF})^\top \mathbf{z}_S^{CF}\right),
\end{align*}
where $\mathbf{W}_s, \mathbf{b}_s$, and $a_s$ denote the weight matrix, bias vector, and activation function for the $s$-th layer's perceptron.
\vspace{-3pt}
\subsection{Drug Collaborative Filtering Module}\label{4.3}
The drug collaborative filtering module performs element-wise product of learned drug representations from the deep network and ADR latent vectors from the shallow network, and then feeds the element-wise product into an FC layer to obtain $\hat y_{i, j}^{DCF}$ as in Figure \ref{fig:1}(d) that
\[\hat y_{i,j}^{DCF}=\sigma\left((\mathbf{h}^{DCF})^\top\left(\phi_L^{D}(\mathbf{z}_{L-1}^{D}) \odot \mathbf{q}_j\right)\right),\]
which has the same form as generalized matrix factorization (GMF)~\cite{he2017neural}, but should be considered as an \emph{asymmetric} collaborative filtering. The key observation is that the learned drug representations are derived from PC-descriptors and BIO-descriptors, which inherently contain extensive structural, chemical, physical and biological features about the drug. Also, the drug representations are fed to an FC layer for drug-ADR prediction in Section \ref{4.1}, so that these representations can retain a strong association to ADR. 
Notably, the ADR latent vectors are \emph{shared} between deep representations and shallow collaborative filtering, which is further interpreted as drug representations can guide the training of ADR latent vectors. We also empirically validate this in our experiments.

\subsection{Joint Training of ADRNet}\label{4.4}
We jointly train the deep drug representation network and the shallow latent collaborative filtering network in ADRNet. Specifically, the drug representation $\phi^D$, its element-wise product with the ADR latent vector $\phi^{DCF}$, and the collaborative filtering vector $\phi^{CF}$ are concatenated and fed into an FC layer for drug-ADR prediction that
\begin{align*}
\phi^{D}_{i, j} & =\phi_L^{D}(\dots(\phi_2^{D}(\phi_1(\mathbf{x}^{PC}_i, \mathbf{x}^{BIO}_i))\dots),\quad
\phi^{CF}_{i, j}  =\phi_S^{CF}(\dots(\phi_2^{CF}(\phi_1(\mathbf{p}_i, \mathbf{q}_j))\dots),\quad
\phi^{DCF}_{i, j} =\phi^{D} \odot \mathbf{q}_j, \\
\hat{y}_{i,j} & =\sigma\left(\mathbf{h}^\top\left[\begin{array}{l}
\phi^{D}_{i, j} \\
\phi^{CF}_{i, j} \\
\phi^{DCF}_{i, j}
\end{array}\right]\right)=\sigma\left(\sigma^{-1}(\hat y_{i, j}^D)+\sigma^{-1}(\hat y_{i, j}^{CF})+\sigma^{-1}(\hat y_{i, j}^{DCF})\right),
\end{align*}
where $\mathbf{h}=[(\mathbf{h}^{D})^\top, (\mathbf{h}^{CF})^\top, (\mathbf{h}^{DCF})^\top]^\top$ denote the edge weights of the output layer. We use mini-batch stochastic optimization to back-propagate the gradients in both the deep and shallow networks, and train the prediction
model by minimizing the training loss
\begin{equation*}
     L(\theta) =-\sum_{i\in\mathcal{I}} \sum_{j\in\mathcal{J}} y_{i,j} \log \hat{y}_{i,j}+\left(1-y_{i,j}\right) \log \left(1-\hat{y}_{i,j}\right).
\end{equation*}
We summarize the advantages of ADRNet over previous works as follows: (a) provides a convenient and flexible framework to combine drug descriptor-based models with latent-based collaborative filtering; (b) has better "memory" and "generalization" ability due to joint learning of shallow and deep models; (c) learns more informative drug and ADR latent vectors compared with pure collaborative filtering.

\section{Experiments}
In this section, we aim to answer the following research questions on real-world clinical and non-clinical datasets:
\begin{itemize}
    \item[$\bullet$] How does the proposed method perform compared with the existing models in terms of drug-ADR prediction? Does non-clinical data help enhance the previous collaborative filtering methods? 
    \item[$\bullet$] What factors (shallow collaborative filtering, deep drug representation, latent embedding, drug descriptor, sharing mechanism) influence the validity of our method?
    \item[$\bullet$] Does our method stably outperform the previous models at varying latent embedding sizes?
    \item[$\bullet$] Are deeper layers in collaborative filtering and drug representation helpful for drug–ADR prediction?
\end{itemize}
	
\subsection{Experimental Setup}
\subsubsection{Dataset and Preprocessing}
We follow the previous studies~\cite{zhang2015predicting,nguyen2021survey} to use two widely used real-world clinical datasets \textsc{Liu's}~\cite{liu2012large} and \textsc{AEOLUS}~\cite{banda2016curated} that contain drug-ADR interactions.  
We then select the drugs that appear in DrugBank, as well as require ADRs to occur on more than 50 drugs. Specifically, the \textsc{Liu's} dataset contains 58,810 drug-ADR interactions occurring on 828 drugs and 1,385 ADRs, and the \textsc{AEOLUS} dataset contains 605,121 interactions occurring on 1,358 drugs and 2,707 ADRs. In addition, to verify the effectiveness of the proposed generalized collaborative filtering framework, we also adopt two real-world non-clinical data PubChem~\cite{kim2016pubchem} and Bio2RDF~\cite{belleau2008bio2rdf}. Specifically, for each drug, PubChem provides pre-defined 881 bits of structural PC-descriptors, and Bio2RDF provides 6,712 bits of chemical, physical, and biological BIO-descriptors. We concatenate the PC-descriptors and BIO-descriptors into 7,593 bits and feed them into the deep drug representation layers.
\vspace{-3pt}
\subsubsection{Baselines}
\vspace{-3pt}
To comprehensively evaluate the proposed method, we compare with the following multi-label prediction methods: \textbf{LNSM}~\cite{zhang2015predicting,munoz2016using}, \textbf{SVM}~\cite{jahid2013ensemble}, \textbf{RF}~\cite{davazdahemami2018chronological,liu2012large}, \textbf{LR}~\cite{cami2011predicting,liu2012large}, \textbf{CCA}~\cite{cami2011predicting,yamanishi2012drug}.
We also compare with the following collaborative filtering methods: \textbf{DrugCF}\footnote{DrugCF is implemented through drug similarity-based collaborative filtering.}, \textbf{MF}~\cite{koren2009matrix,pauwels2011predicting}, 
\textbf{NCF}~\cite{he2017neural},
\textbf{NMF}~\cite{he2017neural}, \textbf{PNN}~\cite{qu2016product}, 
\textbf{FNN}~\cite{zhang2016deep}, 
\textbf{Deep Crossing}~\cite{shan2016deep}, 
\textbf{Wide\&Deep}~\cite{cheng2016wide}, \textbf{Deep\&Cross}~\cite{wang2017deep}, \textbf{DeepFM}~\cite{guo2017deepfm}, \textbf{NFM}~\cite{he2017neural}, \textbf{AFM}~\cite{xiao2017attentional}, and \textbf{UltraGCN}~\cite{mao2021ultragcn}.
\vspace{-3pt}
\subsubsection{Experimental Protocols and Details}
\vspace{-3pt}
\begin{table}
\centering
\footnotesize
\setlength{\tabcolsep}{5pt}
    \caption{Performance comparison of drug–ADR prediction models on \textsc{Liu's}. We bold the best model, and underline the best single-label prediction model and collaborative filtering model.}
    \vspace{-8pt}
    \begin{tabular}{l|cc|r|cc|r}
        \toprule
        Dataset &  \multicolumn{3}{c|}{\textsc{Liu's}} &  \multicolumn{3}{c}{\textsc{AEOLUS}}\\ \midrule
        Method  & 	AUC $(\times 10^{-2})$  & 	AUPR $(\times 10^{-2})$  & Time (s)& 	AUC $(\times 10^{-2})$  & 	AUPR $(\times 10^{-2})$  & Time (s)\\
        \midrule
        LNSM & 90.31 $\pm$ 0.25 &	46.64 $\pm$ 0.69 & 32.6& 84.98 $\pm$ 0.64 &	57.40 $\pm$ 1.63 & 85.7\\
        SVM & \underline{90.98 $\pm$ 0.22} &	\underline{48.79 $\pm$ 0.71} & ~8863.3& \underline{89.25 $\pm$ 0.46} &	\underline{67.54 $\pm$ 1.60} & 20378.0\\
        RF & 87.77 $\pm$ 0.32 &	44.00 $\pm$ 0.59 & 7.7& 	88.56 $\pm$ 0.40 &	66.56 $\pm$ 1.49 & 22.5\\
        LR & 89.25 $\pm$ 0.23 & 46.54 $\pm$ 0.83 & 9.5&  89.02 $\pm$ 1.02 & 66.89 $\pm$ 2.29 & 13.8\\
        CCA & 63.48 $\pm$ 0.83 & 19.50 $\pm$ 0.94 & 72.0& 58.59 $\pm$ 1.51 & 26.69 $\pm$ 1.80 & 141.4\\         \midrule
        DrugCF 
        & 87.88 $\pm$ 0.35 &	45.53 $\pm$ 0.75 & 2.2& 84.79 $\pm$ 0.58 &	58.40 $\pm$ 1.06 & 5.9\\
        MF & 90.89 $\pm$ 0.26 &	45.00 $\pm$ 0.65 & 10.9& 87.15 $\pm$ 0.37 &	61.03 $\pm$ 1.11 & 15.7\\
        NCF &89.84 $\pm$ 0.36 &	44.00 $\pm$ 0.67 & 17.2& 84.68 $\pm$ 0.41	& 59.35 $\pm$ 1.13 & 50.0\\
        NMF & 91.21 $\pm$ 0.37 &	46.73 $\pm$ 0.67 & 19.3& 87.16 $\pm$ 0.32 &	61.03 $\pm$ 1.11 & 43.5\\
        PNN & 91.10 $\pm$ 0.29 &	46.48 $\pm$ 1.06 & 10.1& 90.22 $\pm$ 0.35 &	69.85 $\pm$ 0.96 & 12.7\\
        FNN & 91.05 $\pm$ 0.43 &	45.17 $\pm$ 0.86 & 17.6& 89.64 $\pm$ 0.42 &	66.72 $\pm$ 0.82 & 19.4\\
        Deep Crossing & 90.79 $\pm$ 0.38 &	48.13 $\pm$ 1.09 & 17.2& 88.44 $\pm$ 0.47 &	69.43 $\pm$ 1.23 & 13.5\\
        Wide\&Deep & \underline{91.42 $\pm$ 0.25} & \underline{50.23 $\pm$ 0.85} & 4.6& \underline{90.50 $\pm$ 0.45} & \underline{72.38 $\pm$ 1.11} & 9.2\\
        Deep\&Cross & 90.03 $\pm$ 0.27 & 46.34 $\pm$ 1.12 & 11.7& 87.95 $\pm$ 0.57 & 60.40 $\pm$ 2.91 & 43.2\\
        DeepFM & 91.35 $\pm$ 0.20 & 50.22 $\pm$ 0.75 & 18.9& 90.46 $\pm$ 0.44 & 72.35 $\pm$ 1.04 & 34.2\\
        NFM & 91.41 $\pm$ 0.25 & 50.20 $\pm$ 0.99 & 20.3& 90.40 $\pm$ 0.50 & 71.86 $\pm$ 0.90 & 30.1\\
        AFM & 88.76 $\pm$ 0.53 & 46.05 $\pm$ 1.16 & 3.3& 86.71 $\pm$ 0.57 & 65.63 $\pm$ 1.41 & 5.0\\
        UltraGCN & 91.29 $\pm$ 0.28	& 47.01 $\pm$ 0.69 & 5.0 & 89.54 $\pm$ 0.40	& 71.25 $\pm$ 1.14 & 14.6\\
        \midrule
        ADRNet (ours) &  \textbf{92.23 $\pm$ 0.21*}	& \textbf{51.72 $\pm$ 0.89*} & 5.9&	\textbf{91.20 $\pm$ 0.42*} &	\textbf{74.21 $\pm$ 1.04*} & 15.2     \\
        \bottomrule
    \end{tabular}
    \begin{tablenotes}
\scriptsize
\item Note: * means statistically significant results ($\text{p-value} \leq 0.01$) using the paired-t-test compared with the best baseline.
\end{tablenotes}
    \label{tab:2}
    \vspace{-6pt}
\end{table}

\begin{table*}
    \centering
    \footnotesize
    \setlength{\tabcolsep}{2.5pt}
     \caption{Ablation studies on shallow \emph{versus} deep layers, and latent  collaborative filtering \emph{versus} content-based methods.}
    \vspace{-8pt}
    \begin{tabular}{l|cc|cc|cc|cc}
        \toprule
        & \multicolumn{4}{c|}{Model architecture}  & \multicolumn{2}{c|}{\textsc{Liu's}} & \multicolumn{2}{c}{\textsc{AEOLUS}} \\ \midrule
        Method  & 	Shallow layers  & 	Deep layers	& Latent vector  & Drug descriptor   &	AUC $(\times 10^{-2})$  & 	AUPR $(\times 10^{-2})$  & 	AUC $(\times 10^{-2})$  & 	AUPR $(\times 10^{-2})$\\
        \midrule
        MF     & \checkmark & &  \checkmark& & 90.89 $\pm$ 0.26 &	45.00 $\pm$ 0.65 &	87.15 $\pm$ 0.37 &	61.03 $\pm$ 1.11        \\
        NCF   &  & \checkmark &  \checkmark& &89.84 $\pm$ 0.36 &	44.00 $\pm$ 0.67 &	84.68 $\pm$ 0.41	& 59.35 $\pm$ 1.13        \\
        NMF & \checkmark & \checkmark  &\checkmark& &91.21 $\pm$ 0.37 &	46.73 $\pm$ 0.67 &	87.16 $\pm$ 0.32 &	61.03 $\pm$ 1.11        \\ \midrule
       LR  & \checkmark &  & &\checkmark  &89.25 $\pm$ 0.23 &	46.54 $\pm$ 0.83 &	89.02 $\pm$ 1.02 &	66.89  $\pm$  2.29        \\
         MLP  &  & \checkmark & &\checkmark  &90.23 $\pm$ 0.29 &	45.64 $\pm$ 0.79 &	89.48 $\pm$ 0.08 &	68.65 $\pm$ 1.19        \\
        Wide\&Deep & \checkmark & \checkmark & &\checkmark& 91.42 $\pm$ 0.25 &	50.23 $\pm$ 0.85 & 90.50 $\pm$ 0.45 &	72.38 $\pm$ 1.11        \\
        \midrule 
        ADRNet w/o share & \checkmark & \checkmark & \checkmark& \checkmark & \underline{92.16 $\pm$ 0.21*} &	\underline{51.71 $\pm$ 0.93*} &	\underline{91.14 $\pm$ 0.40*} &	\underline{73.91 $\pm$ 1.09*}     \\
        ADRNet & \checkmark &\checkmark&  \checkmark & \checkmark & \textbf{92.23 $\pm$ 0.21*}	& \textbf{51.72 $\pm$ 0.89*} &	\textbf{91.20 $\pm$ 0.42*} &	\textbf{74.21 $\pm$ 1.04*}     \\
        \bottomrule
    \end{tabular}
    \begin{tablenotes}
\scriptsize
\item Note: * means statistically significant results ($\text{p-value} \leq 0.01$) using the paired-t-test compared with the best baseline.
\end{tablenotes}
    \label{tab:4}
    \vspace{-12pt}
\end{table*}
Following previous studies~\cite{yamanishi2012drug,zhang2015predicting,pauwels2011predicting,dimitri2017drugclust}, we use 10-fold cross-validation for our experiments. Two common metrics are used to evaluate the prediction performance, i.e., area under the ROC curve (AUC) and area under the precision-recall curve (AUPR). All the experiments are implemented on PyTorch with Adam as the optimizer
. We tune the learning rate in $\{0.001, 0.005, 0.01, 0.05 \}$, weight decay in $\{1e-6, 1e-5, 1e-4, 1e-3\}$, and embedding size in $\{16, 32, 64, 128, 256, 512, 1024\}$ for both \textsc{Liu's} and \textsc{AEOLUS}.

\subsection{Performance Comparison}

\begin{figure}[t]
\centering
\resizebox{1\linewidth}{!}{
\subfigure[AUC on \textsc{Liu's}]{
\begin{minipage}[t]{0.3\linewidth}
\centering
\includegraphics[width=1\textwidth]{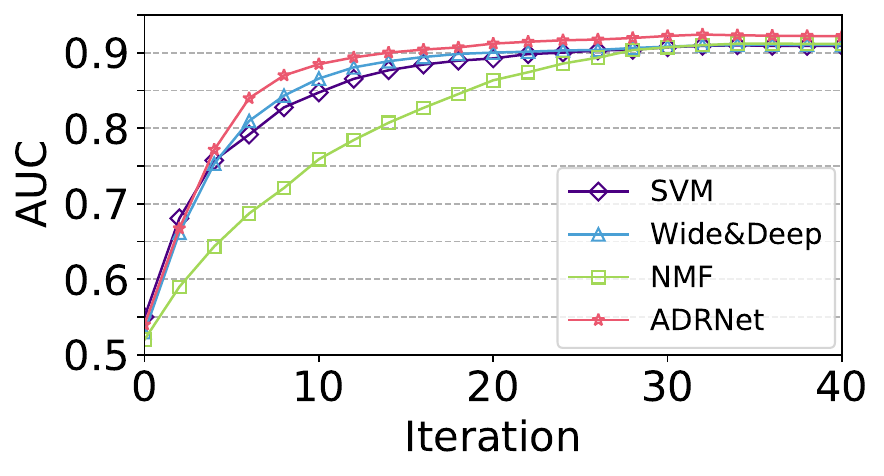}
\end{minipage}%
}%
\subfigure[AUPR on \textsc{Liu's}]{
\begin{minipage}[t]{0.3\linewidth}
\centering
\includegraphics[width=1\textwidth]{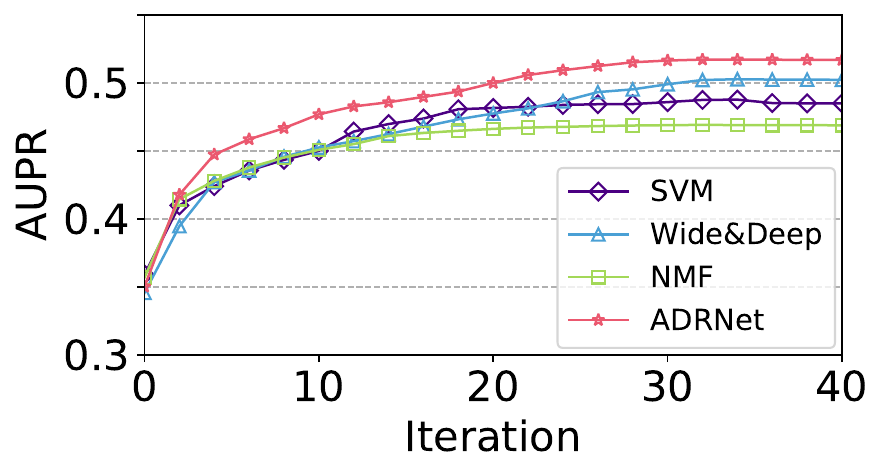}
\end{minipage}%
}
\subfigure[AUC on \textsc{AEOLUS}]{
\begin{minipage}[t]{0.3\linewidth}
\centering
\includegraphics[width=1\textwidth]{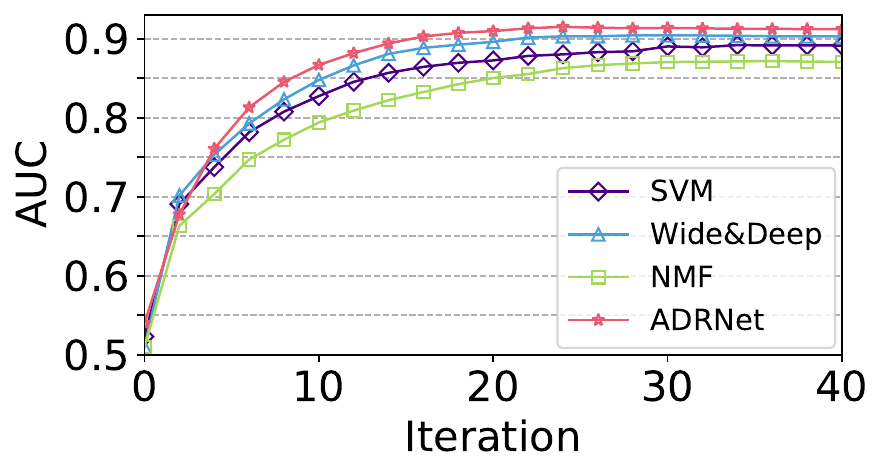}
\end{minipage}%
}%
\subfigure[AUPR on \textsc{AEOLUS}]{
\begin{minipage}[t]{0.3\linewidth}
\centering
\includegraphics[width=1\textwidth]{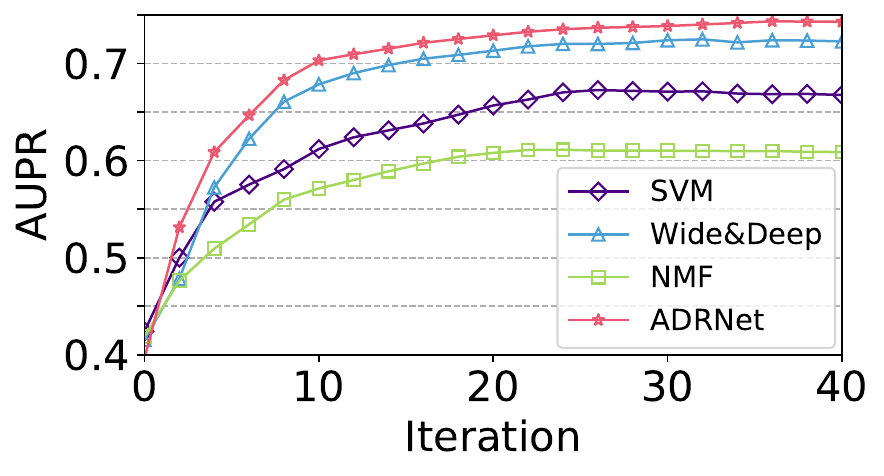}
\end{minipage}%
}}
\centering
\vspace{-12pt}
\caption{Prediction performance of the ADRNet, compared with the best single-label prediction model (SVM), multi-label prediction model (Wide\&Deep), latent collabrative filtering model (NMF), w.r.t. the number of iterations on \textsc{Liu's} and \textsc{AEOLUS}.}
\vspace{-12pt}
\label{fig:2}
\end{figure}

\emph{Overall Performance.} 
We compare the proposed ADRNet with a wide range of baseline methods on two real-world datasets \textsc{Liu's} and \textsc{AEOLUS}, and the results are shown in Table \ref{tab:2}. 
We find the proposed ADRNet outperforms the best baseline in terms of AUC and AUPR at 0.01 statistical significance. We also compare the running times of the various methods in Table \ref{tab:2}. Compared with the content-based collaborative filtering, ADRNet has a lower running time due to the faster convergence resulting from the joint training of the deep network with latent collaborative filtering.
\vspace{4pt}\\
\emph{Analysis of Convergence.} To further explore the convergence performance of ADRNet, we compare 
AUCs and AUPRs with respect to the number of iterations in Figure \ref{fig:2}. We find that ADRNet using joint training has the fastest convergence speed on both datasets, while NMF converges the slowest. This indicates the guiding effect of drug descriptor-based representations on the latent vectors in collaborative filtering.\vspace{0.25pt}
\subsection{Ablation Studies}
\begin{figure}[t]
\centering
\resizebox{1\linewidth}{!}{
\subfigure[AUC on \textsc{Liu's}]{
\begin{minipage}[t]{0.3\linewidth}
\centering
\includegraphics[width=1\textwidth]{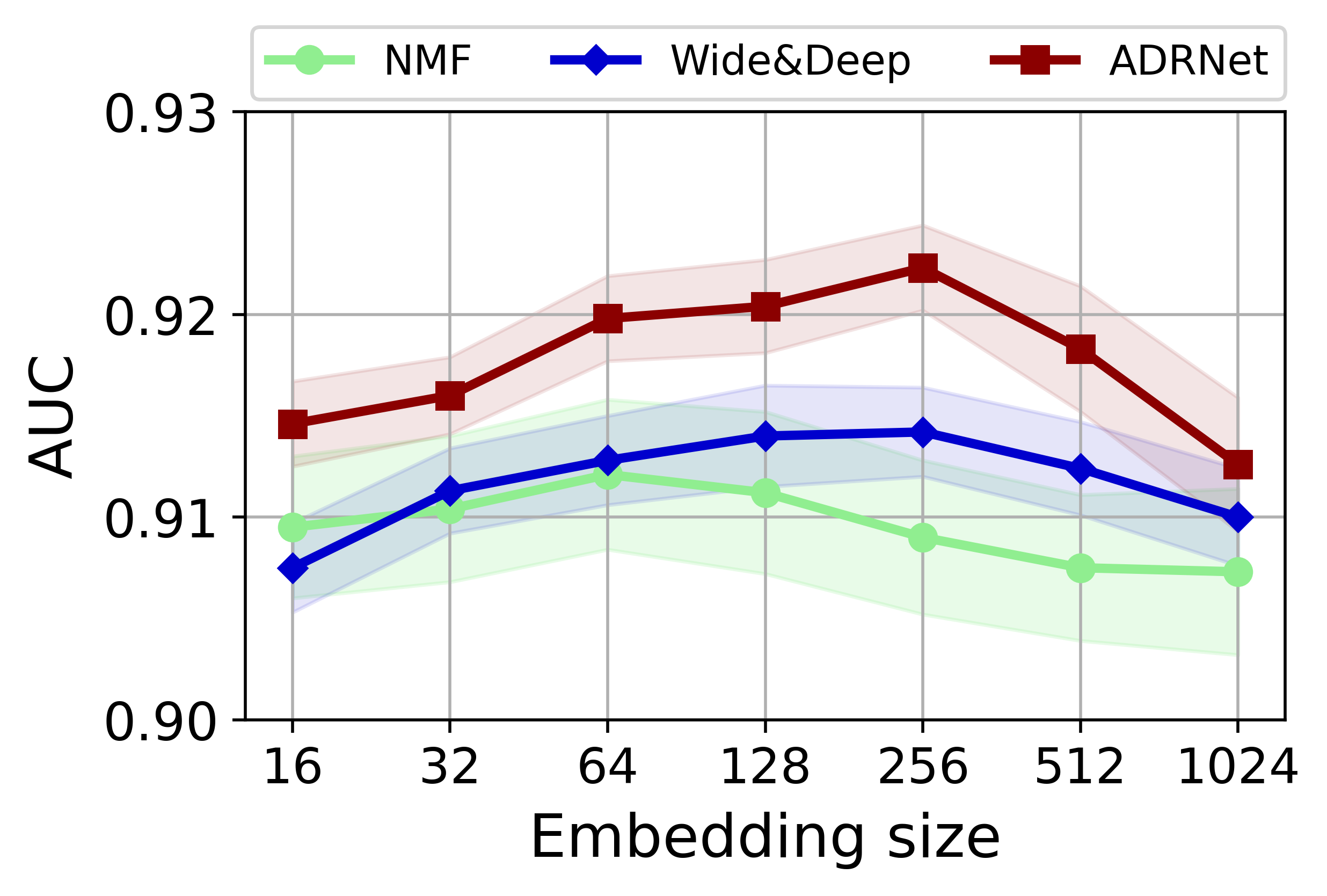}
\end{minipage}%
}%
\subfigure[AUPR on \textsc{Liu's}]{
\begin{minipage}[t]{0.3\linewidth}
\centering
\includegraphics[width=1\textwidth]{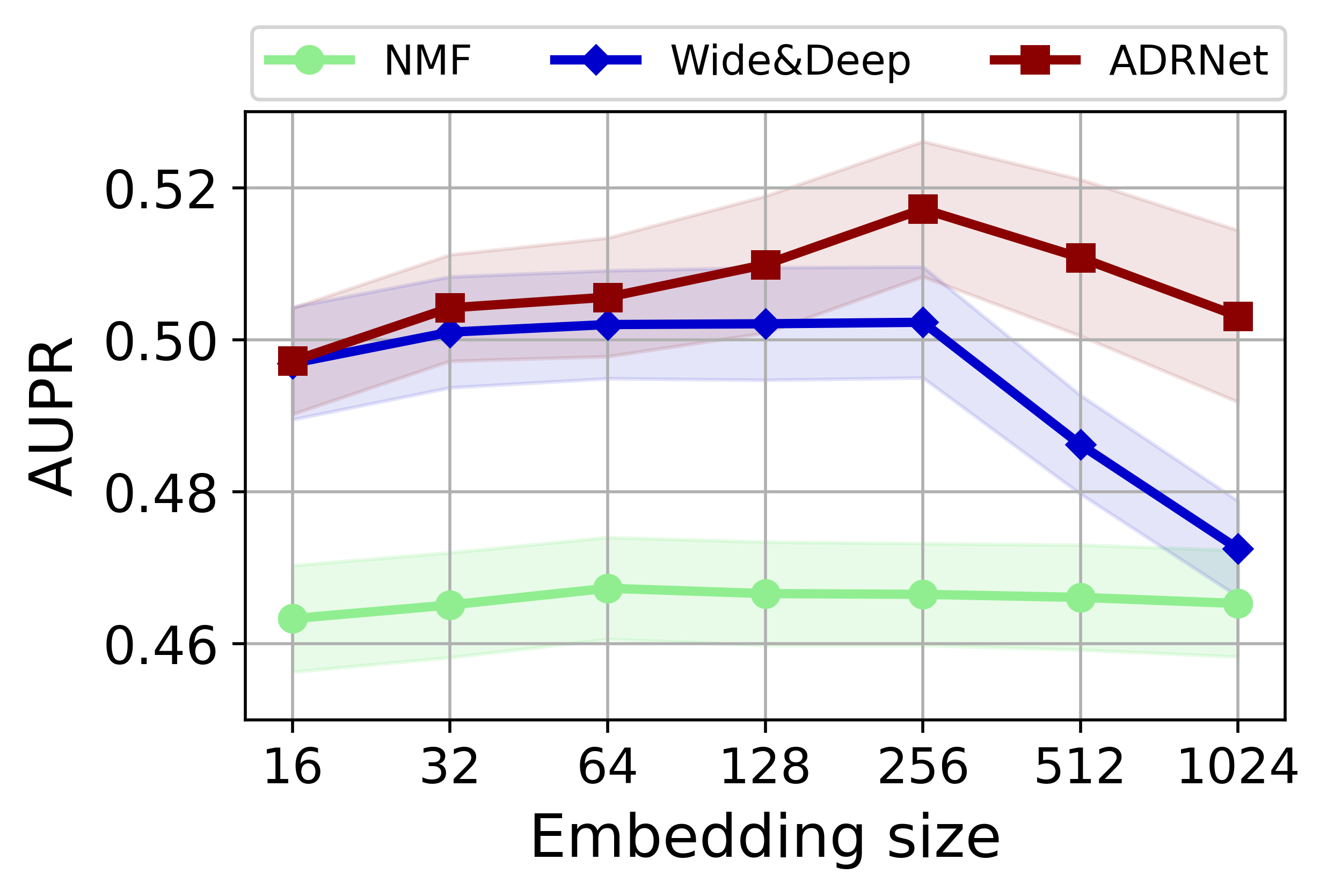}
\end{minipage}%
}%
\subfigure[AUC on \textsc{AEOLUS}]{
\begin{minipage}[t]{0.3\linewidth}
\centering
\includegraphics[width=1\textwidth]{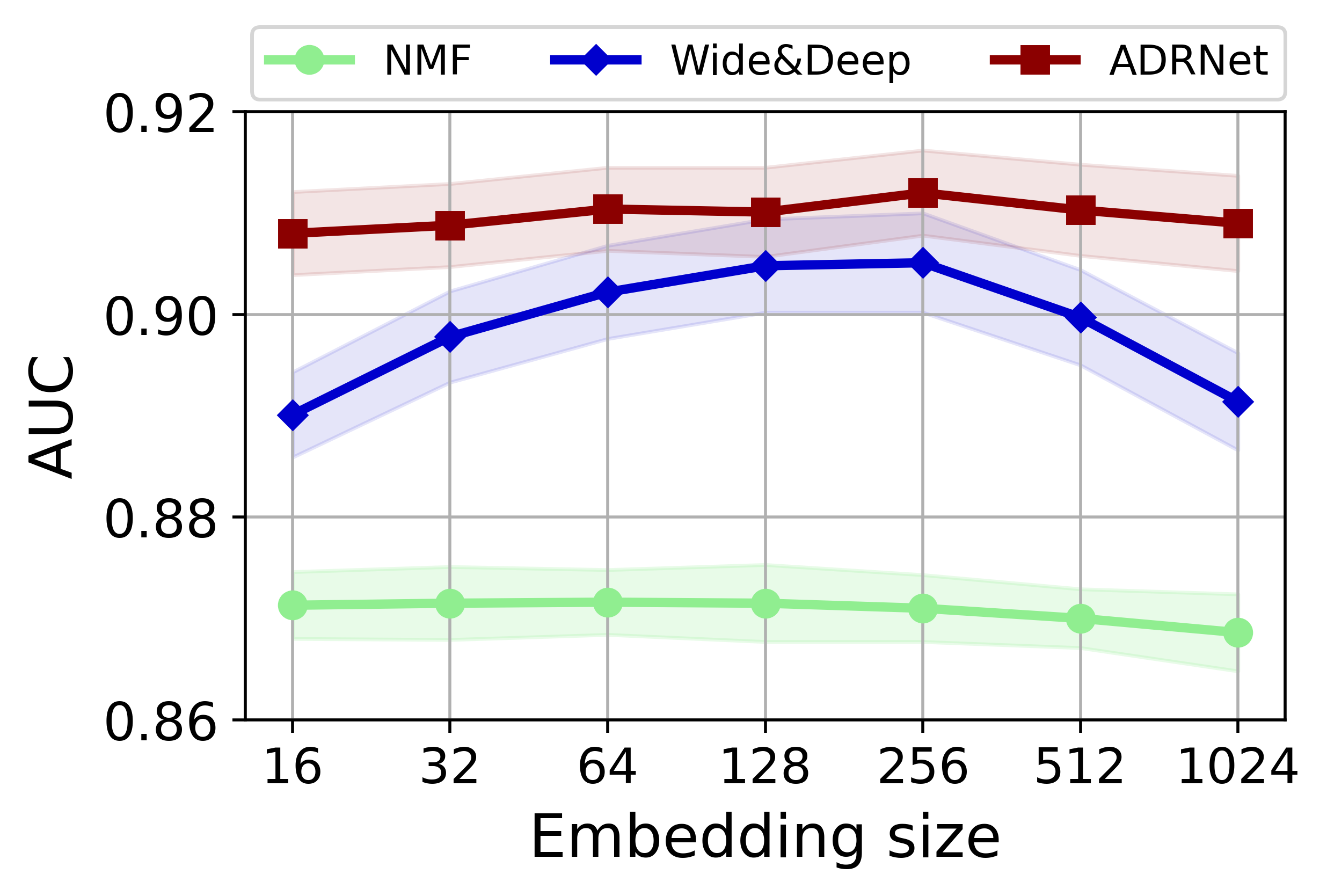}
\end{minipage}%
}%
\subfigure[AUPR on \textsc{AEOLUS}]{
\begin{minipage}[t]{0.3\linewidth}
\centering
\includegraphics[width=1\textwidth]{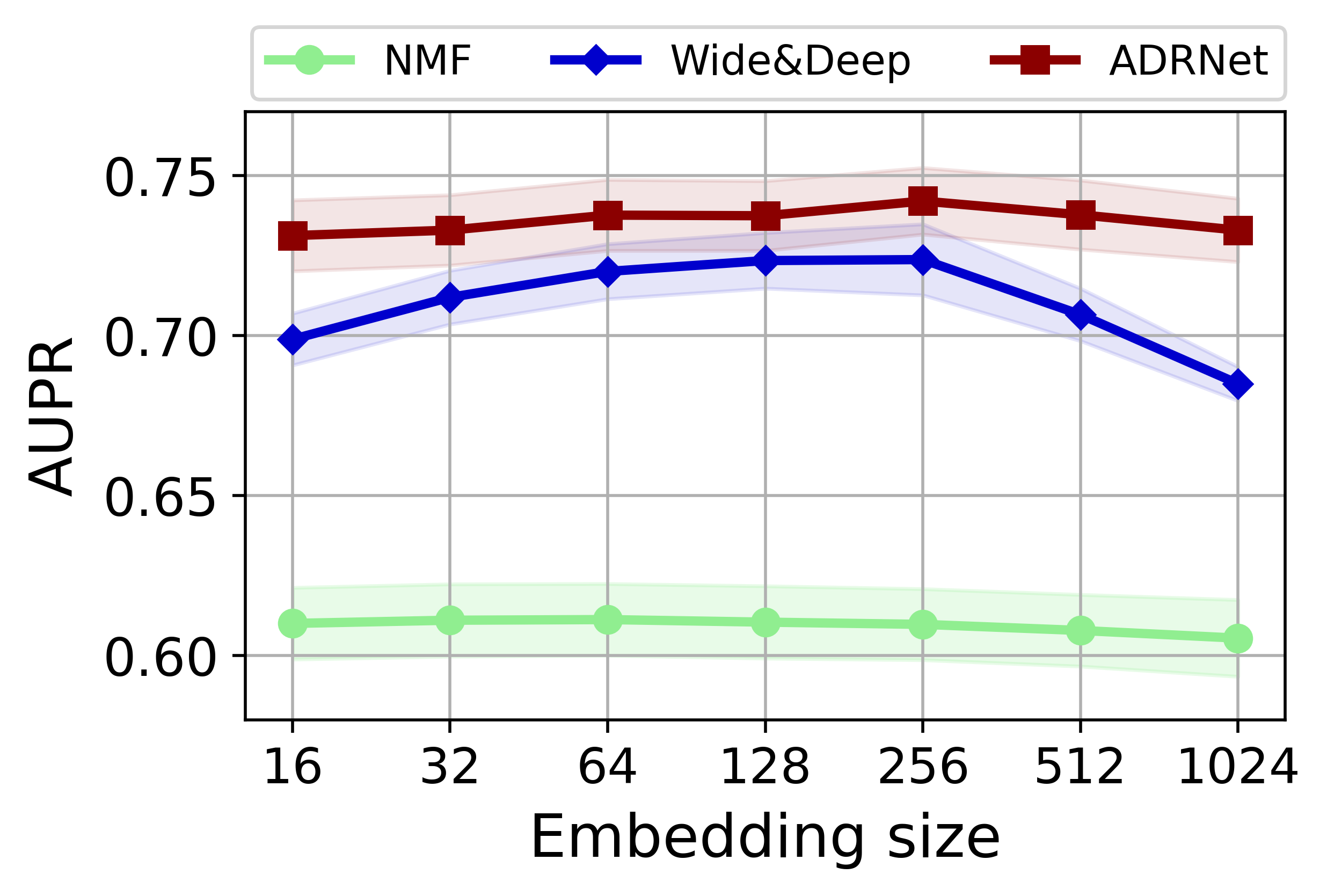}
\end{minipage}%
}}%
\centering
\vspace{-12pt}
\caption{Effects of the embedding size on \textsc{Liu's} and \textsc{AEOLUS}.}
\vspace{-2pt}
\label{fig:3}
\vspace{-10pt}
\end{figure}

\vspace{2pt}
\emph{Shallow vs. Deep Layers.} ADRNet has both shallow latent collaborative filtering and deep drug representation layers. In Table \ref{tab:4}, we conduct ablation experiments to explore the effects of using the shallow network, using the deep network, and using both on drug-ADR prediction performance. We find that using both shallow and deep networks helps to improve the prediction performance in terms of AUC and AUPR, which is consistent with the discussion in Section \ref{4.2}. 
\vspace{5pt}\\
\emph{Latent Vector vs. Drug Descriptor.}  We also explore the effect of latent vectors versus drug descriptors on prediction performance, and the results are presented in Table \ref{tab:4}. We find that the high-dimensional drug descriptor-based models outperform the pure collaborative filtering models significantly. This is due to the structural, chemical, physical, and biological information about the drug contained in the drug descriptors. 
\vspace{5pt}\\
\emph{Embedding Size.} We next investigate how embedding size affects drug-ADR prediction performance. Specifically, we compare ADRNet with the best pure (NMF) and the best content-based (Wide\&Deep) collaborative filtering that uses both shallow and deep networks, with the results shown in Figure \ref{fig:3}. Notably, ADRNet stably outperforms NMF and Wide\&Deep at all embedding sizes and achieves the optimal performance when embedding sizes are around 128-256. 

\section{Conclusion}
We formulate the prediction of multi-label ADRs as a drug-ADR collaborative filtering problem, and to the best of our knowledge, this is the first study to provide extensive benchmark results of previous collaborative filtering methods on two large publicly available clinical datasets. Then, by noting the gap between previous studies on collaborative filtering with symmetric information and the asymmetry of available drug and ADR information for multi-label ADR predictions, we propose ADRNet as a generalized collaborative filtering framework combining clinical and non-clinical data for drug-ADR prediction. We describe and discuss in detail the three modules of ADRNet: a deep drug representation module, a shallow latent collaborative filtering module, and a drug collaborative filtering module. Notably, ADRNet inherits the "generalization" ability of the deep model and the "memory" ability of the latent collaborative filtering, and further shares the ADR latent vector to establish stronger connection. Finally, we jointly train the deep drug representation and the shallow collaborative filtering network to better trade-off the predictions of each sub-network. Extensive experiments are conducted on two publicly available real-world drug-ADR clinical datasets and two non-clinical datas, and the results show that the proposed ADRNet can effectively combine the advantages of drug descriptors and latent collaborative filtering to improve drug-ADR prediction performance.


\bibliographystyle{ACM-Reference-Format}
\bibliography{abbrev}




\end{document}